\documentclass[twocolumn,pre,showpacs]{revtex4}
\usepackage{graphicx}
\usepackage{dcolumn}
\usepackage{amsmath}
\usepackage{latexsym}

\def\bea{\begin{eqnarray}}
\def\eea{\end{eqnarray}}
\def\beq{\begin{equation}}
\def\eeq{\end{equation}}

\def\D{\Delta}

\def\d{\delta}

\def\o{\omega}

\def\d{\delta}

\def\o{\omega}

\begin{document}
\title{Anderson orthogonality catastrophe in realistic quantum dots}
\author{Swarnali Bandopadhyay}
\affiliation{Department of Physics, Norwegian University of Science and Technology, N-7491, Trondheim, Norway}
\email{swarnali.bandopadhyay@ntnu.no}
\author{Martina Hentschel}
%\email{martina@mpipks-dresden.mpg.de}
\affiliation{
Max-Planck-Institut f\"ur Physik komplexer Systeme,
%Max Planck Institute for the Physics of Complex Systems, 
N{\"o}thnitzer Str. 38, 01187 Dresden, Germany
}
\date{\today}

\begin{abstract}
We study Anderson orthogonality catastrophe (AOC) for an   
parabolic quantum dot (PQD), one of the experimentally
realizable few-electron systems. 
The finite number of electrons in PQD causes AOC 
to be incomplete, with a broad 
distribution of many-body overlaps. 
This is a signature of mesoscopic fluctuations
and is in agreement with earlier results obtained for chaotic quantum dots.
Here, we focus on the effects of degeneracies in  PQDs, 
realized through their inherent shell structures, 
on AOC. We find rich and interesting behaviours as a 
function of the strength and position of the perturbation, the system size, 
and the applied magnetic field. In particular, even for weak perturbations, 
we observe a pronounced AOC which is related to the degeneracy of energy 
levels. Most importantly, the power law decay 
of the many-body overlap as a function of increasing number of particles
is modified in comparison to the metallic case due to
rearrangements of energy levels in different shells.   

\end{abstract}
\maketitle
%%%%%%%%%%%%%%%%%%%%%%%%%%%%%%%%%%%%%%%%%%%%%%%%%%%%%%%%%%%%%%%%%%%%%%%%%%
\section{Introduction}\label{intro}
%%%%%%%%%%%%%%%%%%%%%%%%%%%%%%%%%%%%%%%%%%%%%%%%%%%%%%%%%%%%%%%%%%%%%%%%%%
Anderson orthogonality catastrophe (AOC) is one of the simplest many-body 
effects in condensed matter physics. It was first described by 
Anderson~\cite{anderson} in 1967 and refers to the vanishing of the 
overlap between the unperturbed and perturbed many-body ground states as 
a power law in the number of particles  
in the system. This 
results from the non-adiabatic response of the system to a sudden perturbation. 
AOC contributes to a number of Fermi-edge singularities (FES), e.g.,  
in the Kondo effect~\cite{hewson} or in the X-ray edge 
problem~\cite{mahan,nozieres, tanabe}. In the X-ray edge problem, the sudden 
and localised perturbation is realized through the X-ray excitation of a core 
electron that leaves behind a localised attracting (hole) potential. Here, 
other many-body responses may play a role, e.g.,  
the so-called Mahan-Nozi\`eres-DeDominicis response or Mahan's 
exciton~\cite{mahan,nozieres} in FES of photoabsorption spectra. 
However, in the case of photo-emission experiments where the excited 
core electron leaves the sample, the physics is governed by  AOC 
alone~\cite{kroha}. This will be our focus in this article.

AOC was originally introduced and discussed for 
bulk samples with a large number of particles (corresponding to the 
thermodynamic limit). The conduction electrons respond to the attractive 
core-hole potential by slightly lowering their single particle energy levels. 
Although the overlap between the single particle states before and after the
 photo-excitation remains very close to one, this is not true for the overlap 
between the corresponding many-body ground states. In  AOC 
the many-body overlap between the unperturbed and perturbed ground 
states $|\Delta|^2$ vanishes as a power law in the number of participating 
electrons $M_{\mbox{eff}}$ as
$$ |\Delta|^2 \propto M_{\mbox{eff}}^{-\phi^2}\,\,, $$
with $M_{\mbox{eff}}=\rho_c W_c$ where $\rho_c$ is the density of states of the
conduction band at the Fermi level and $W_c$ is the width of the conduction 
band \cite{tanabe}; $\phi$ is the phase shift at the Fermi energy. 

In recent years, the fabrication of high-quality samples with a finite 
number of electrons ranging from very few to, say, several thousands,  
has become possible. A hallmark example are the effectively 
two-dimensional (2D) quantum dots realized in  semiconductor 
heterostructures ~\cite{qdref} where electrons occupy well defined discrete 
levels that can be probed by Coulomb-blockade measurements \cite{sohn_etal}. 
The size of the sample can even be made smaller than the phase-coherence 
length of the system. These samples are referred to as ballistic. 
Consequently, in the spirit of quantum chaos, the geometry of 
the quantum dot becomes important as the wavefunctions and energy 
eigenstates  depend sensitively on the geometry of the system's 
boundary due to self-interference effects~\cite{stoeckmann,marcus}. This 
new aspect, besides the much smaller number of 
electrons in the conduction band, has to be contrasted to Bloch waves in 
metals. 
The existence of mesoscopic fluctuations is another feature 
that governs the behaviour of such systems. 
All these differences from bulk systems  
have made mesoscopic systems an interesting object of research in the context 
of AOC and  FES ~\cite{gefen1,gefen2,hentschel1,hentschel2,hentschel3}. 

\begin{figure}[]
\includegraphics[width=6.5cm]{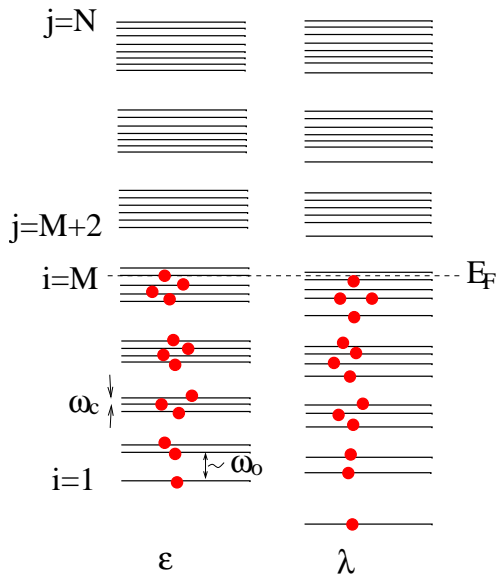}
\caption{A schematic diagram of the
energy levels  ($\varepsilon$) in a PQD in the presence of a small 
magnetic field (characterised by cyclotron frequency $\omega_c$) that slightly 
lifts the degeneracy of the PQD's energy-levels.  
Notice the shell structure. An attractive
perturbation ($V_c$) due to a core hole shifts the levels  
downwards and results in perturbed levels $\lambda$. $M$ is the 
number of occupied dot levels (to the Fermi energy $E_F$) and $N$ is 
the total number of levels.
}
\label{PQD}
\end{figure}

Here, we focus in particular on yet another difference from the bulk case that can easily 
be realized in mesoscopic structures, namely, the existence of degenerate 
energy levels (shells): Quantum dots with a few up to few hundred electrons 
are often very well described by a parabolic confining 
potential~\cite{kouwenhoven} that leads, in close analogy to the harmonic 
oscillator, to the organisation of the energy levels in shells (see 
Fig.~\ref{PQD}). The objective 
of the present paper is to study AOC in the presence of degeneracies and 
shell structures as they occur in realistic, parabolic quantum dots where, 
furthermore, an external magnetic field can be used to control the 
lifting of degeneracies.
We show that the presence of shells considerably alters AOC of 
the system and leads to deviations from the power-law behaviour 
observed in the bulk systems.

The paper is organised as follows. In Sec.~\ref{model} we describe our 
model that follows the Fermi golden 
rule approach to FES in the x-ray edge problem introduced by Ohtaka and Tanabe in Refs.~\cite{tanabe,rank1}. The sudden, local perturbation 
arising from the core hole is modelled as a rank-one perturbation. In 
Sec.~\ref{result} we present our results for AOC in parabolic quantum dots. 
First, we focus on the role of shell effects and neglect the dependence of the 
perturbation strength on the position of the perturbation. We then turn to the 
more realistic case where the dependence of the perturbation strength on the 
position, induced by the non-homogeneous electron intensity in PQD, as well as 
mesoscopic fluctuations are taken into account. For each 
case we consider both weak and strong magnitude of attractive perturbation
and a weak magnetic control. Finally we draw conclusions in Sec.~\ref{concl}.

%%%%%%%%%%%%%      end of Introduction      %%%%%%%%%%%%%%%%%%%%%

%--------------------------------

%%%%%%%%%%%%%%%%%%%%%%%%%%%%%%%%%%%%%%%%%%%%%%%%%%
\section{The model }\label{model}
%%%%%%%%%%%%%%%%%%%%%%%%%%%%%%%%%%%%%%%%%%%%%%%%%%
We start with a discussion of the energy levels and wavefunctions of the 
(unperturbed) PQD, paying special attention to the degeneracy of levels and 
their organisation in shells as this has crucial impact on the Anderson 
overlap. Its calculation is discussed in the second subsection. 
\subsection{Energy levels and wave functions of the parabolic quantum dot}
We describe the unperturbed system by the Hamiltonian
\begin{equation}
\hat H_0 =\sum_{k=0}^{N-1}\, \epsilon_k\,c_k^\dagger\,c_k
\end{equation}
where the operator $c_k^\dagger$ ($c_k$) creates (annihilates) a particle 
in the unperturbed eigenstate $\psi_k(\vec r)$ with eigenenergy $\epsilon_k$. 

\begin{figure}[tb]
\includegraphics[width=8cm]{unpertEngatwcinset.eps}
\includegraphics[width=9.5cm]{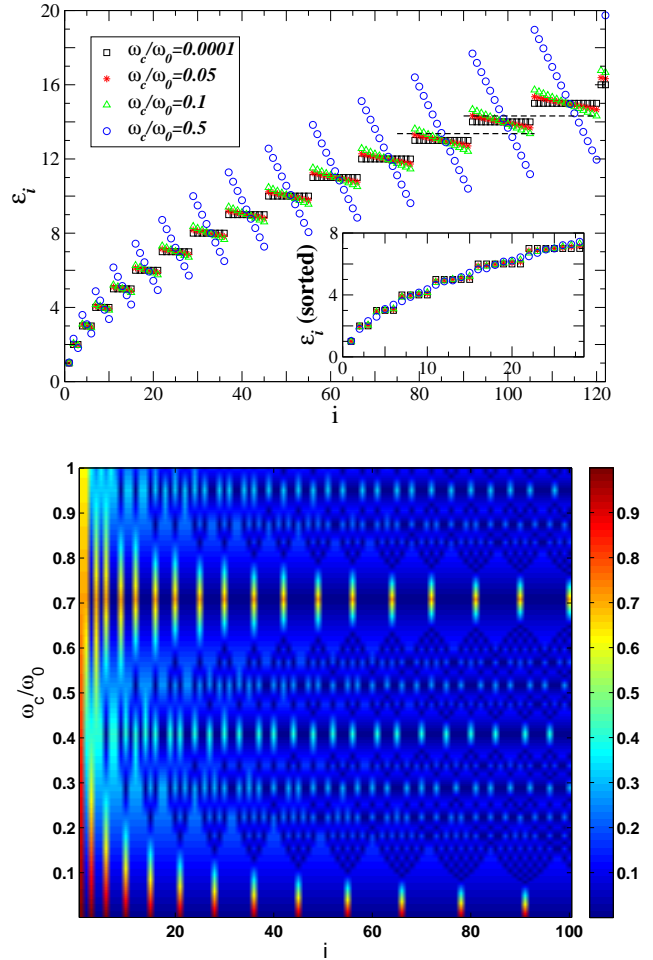}
\caption{(Color online) Upper panel: Energy $\epsilon_i$ of the first 
$i=1, 2, \cdots 120$ unperturbed levels of a PQD with fixed total number of 
levels $N=250$ and fixed $\omega_0$=1 for four different $\omega_c$.
 The shell (``cluster'') structure is clearly visible. For higher
shells, mixing between adjacent clusters starts at lower $\omega_c$. Two 
dashed lines are shown as guide for the eye to indicate mixing 
at $\omega_c/\o_0=0.1$. The inset shows the energy-sorted levels. 
Lower panel: The separation between two adjacent levels 
$\delta \epsilon_i=\epsilon_{i+1}-\epsilon_i$ of the same PQD is shown as
a color %contour 
plot as a function 
of magnetic control $\omega_c/\omega_0$ and energy level $i$. Note that a new shell structure forms near $\o_c/\o_0=0.7$.
}
\label{engsp}
\end{figure}

A very weak magnetic field will be applied to lift the inherent degeneracy 
in the PQD energy levels. Tuning the strength of the applied field, 
characterised by the cyclotron frequency $\omega_c$, we can study 
quasi-degenerate to non-degenerate limits. In presence of a magnetic 
field, the eigenfunction and the corresponding eigenenergies of the PQD 
are \cite{kouwenhoven}
\begin{widetext}
\begin{eqnarray}
\psi_{n,l}(r,\phi)&=&\frac{e^{il\phi}}{\sqrt{2\pi}l_b}\,\sqrt{\frac{n!}{(n+|l|)!}}\,\exp(-\frac{r^2}{4l_b^2})\,
\left(\frac{r}{\sqrt{2}l_b}\right)^{|l|}\,L^{|l|}(\frac{r^2}{2\,l_b^2})\label{wavefn} \\
\epsilon_{n,l}&=&(2n+|l|+1)\,\hbar\,(\omega_0^2+\frac{1}{4}\,\omega_c^2)^{1/2}-\frac{1}{2}\,l\hbar\,\omega_c\label{eigE}\,,
\end{eqnarray}
\end{widetext}
where $\omega_0$ is the oscillator frequency ($\omega_0=3meV$ for GaAs) and
$\omega_c=\frac{\hbar e B}{m^*}$ is the cyclotron frequency (with the effective mass $m^*=0.067\,m_e$ for GaAs).
The resulting characteristic frequency of the oscillator is $\Omega=
(\omega_0^2+\frac{1}{4}\,\omega_c^2)^{1/2}$. 
Furthermore, $n(=0,1,2,\cdots \cdots)$ 
is the radial quantum number and $l(=0,\pm 1,\pm 2,\cdots \cdots)$
is the angular momentum quantum number. $l_b(=\sqrt{\frac{\hbar}{m^*\Omega}})$ 
is the characteristic length of the oscillator.  
$L_n^{|l|}(x)=\sum_{i=0}^n\,(-x)^i\,\frac{(n+|l|)!}{(n-i)!\,(|l|+i)!\, i!}$ 
is the generalised (associated) Laguerre polynomial.
We use units $\hbar=1, m^*=1, \omega_0=1$ in the following. 
We will consider PQDs with a total number of shells (``clusters'') 
$S(=2n+|l|+1)=1,2,\cdots \cdots 50$. Neglecting spin degrees of
freedom, the $p$-th cluster contains $p$ levels (Fig.~\ref{engsp}).

The energy gap between two adjacent clusters is defined as the difference between 
the top-most level of the lower ($i$-th) cluster and the lowest level of next higher 
($i+1$)-th cluster, i.e.,
\bea
\Delta \epsilon&=&[(i+1)-i]\Omega \,-\,
\frac{1}{2}\,\left(\left|l_{\mbox{max}}^{(i+1)}\right|\,+\,\left|l_{\mbox{max}}^{(i)}\right|\right)\,\omega_c\label{Eshi}%\\
\eea
Using $\left|l_{\mbox{max}}^{(i)}\right|=(i-1)$ in Eq.~(\ref{Eshi}),
we find $\Delta \epsilon= \Omega-(i-1/2)\,\omega_c$. 
Thus for given $\omega_0$ and $\omega_c$ with $\omega_c<<\omega_0$, $\Delta \epsilon$ 
decreases for higher clusters.

%-------------------------------------
%---------------------------------

Figure \ref{engsp} shows the unperturbed levels as expressed in Eq.~(\ref{eigE}) 
for few values of $\o_c/\o_0$ (upper panel). For small $\o_c/\o_0$,  i.e., in the 
quasi-degenerate limit all the clusters are well-separated from their neighbours. 
With increasing $\o_c/\o_0$, levels 
from adjacent clusters start to mix in, eventually destroying the shell structure. 
For higher (and therefore larger) clusters, mixing starts for comparatively 
weaker $\omega_c$ than for smaller clusters. For example, in Fig.~\ref{engsp}, 
for $\omega_c=0.1\omega_0$ there is no mixing up to the tenth cluster, 
whereas for $\omega_c=0.5\omega_0$ mixing starts from third cluster. 
In the inset, we have plotted the sorted unperturbed levels of our system. 
Clustering of levels at weak $\omega_c$ fades away with the increasing  
magnetic control parameter.  For $\omega_c=0.5\omega_0$, levels are almost 
equi-spaced.  In the lower panel of Fig.~\ref{engsp}, we have shown the 
energy difference of adjacent levels as a function of level 
index and magnetic control parameter. For small $\omega_c$ ($\simeq 0.1\o_0$), 
in the absence of mixing, the energy separation between two adjacent 
clusters is constant and given by  $\Delta \epsilon \sim \o_0$ with 
an intra-cluster level spacing of $\Delta_{ic} \epsilon \sim \o_c$. 
With increasing $\omega_c$, as levels from different clusters start 
to mix in, the energy-spacing becomes a complicated function 
of  $\omega_0$, $\omega_c$ and the cluster size. For certain ranges 
of $\omega_c/\o_0$ (e.g., around $\o_c/\o_0=0.7$) one obtains 
an almost uniform level spacing. If one increases  $\omega_c$ further,  
levels start to form clusters again. However, these clusters are very 
different from the initial shell structure.

In the absence of a magnetic field, one can estimate the effective 
radius $r_{\mbox{eff}}$ of the PQD 
by equating the energy $\frac{1}{2}\,m^*\,\omega_0^2\,r_{\mbox{eff}}^2$ 
at the classical turning point to the energy of the highest filled 
shell~\cite{kouwenhoven}. For a PQD with $15$ shells and
half-filling, i.e., a partially filled eleventh shell, corresponding 
to an energy $\sim 11\,\hbar\,\omega_0$, the effective radius is 
$r_{\mbox{eff}}=\sqrt{2\,\times\,11}$ (which is $4.7\,\times 10^{-7}\,m$
 for the above-mentioned GaAs quantum dot). 

\subsection{Calculation of the Anderson overlap}

Following the approach by Tanabe and Ohtaka~\cite{tanabe,rank1}, we model the localised 
perturbation, generated by the sudden appearance of the core hole left behind 
after x-ray excitation of one core electron, as a local potential described as 
rank-one perturbation. For details of the underlying theory we refer the reader, e.g.,  
to Refs.~\cite{tanabe}, \cite{hentschel2} or \cite{rank1}. The great advantage of a rank-one 
perturbation is that all quantities of interest can be expressed in terms of the 
unperturbed and perturbed energy levels (which, however, depend on the wave-function 
amplitude at the position of the perturbation; see Eq.(~\ref{OvorgF}) in the appendix). 
Nevertheless it was found to provide a very reasonable description of the 
many-body effects contributing to the x-ray edge problem \cite{tanabe}.

We write the rank-one contact potential as
\begin{equation}
\hat V_c =V_c \Omega_D f^\dagger(\vec r_c)\,f(\vec r_c)\,,
\end{equation}
with $f(\vec r_c)=\sum_k\,\psi_k(\vec r_c)\,c_k$, that acts only at the location
 $\vec r_c$ of the core hole. $\Omega_D$ is the volume of the PQD 
 and the parameter $V_c$ defines the strength of the potential. Note that 
its effective strength depends also on $\psi_k (\vec r_c)$, i.e., on the amplitude 
of the wave function at the position of the perturbation. 
A small wave-function amplitude at $\vec r_c$ will reduce the effective 
perturbation strength felt by the system.
 The energy levels $\epsilon$ will move under the influence of $\hat{V}_c$,
downward for an attractive potential (core hole) and upward for a repulsive one.
We focus on an attractive perturbation $V_c<0$ (core hole) in what follows. 
$V_c$ introduces a new energy scale into the problem, and 
$V_c/\omega_0$ and $\omega_c/\omega_0$ are the two
dimensionless energy-scales of the PQD that govern the system's behaviour. 

Introducing $d_{\kappa}^\dagger$ as creation operator of a particle in the perturbed orbital 
$\phi_{\kappa}$, we can write the perturbed Hamiltonian in diagonal form as
\begin{equation}
\hat H = \hat H_0 + \hat V_c =\sum_{\kappa}\, \lambda_{\kappa}\,d_{\kappa}^\dagger\,d_{\kappa}\,.
\label{Htot}
\end{equation}
To obtain the perturbed single particle states $\phi_{\kappa}$ and
 their eigenvalues $\lambda_{\kappa}$, we diagonalise the total Hamiltonian
$\hat H$. 
The unperturbed and perturbed many-body ground states
for $M$ electrons are obtained as Slater determinants,
$\Psi_0=|\psi_1,\psi_2,\cdots ,\psi_M|$ (unperturbed)
and $\Phi_0=|\phi_1,\phi_2,\cdots ,\phi_M|$ (perturbed). 
The Anderson overlap of these two many-body states is given by
$|\Delta|^2=|\langle\Phi_0| \Psi_0\rangle|^2$, with $\Delta=det(A)$.
The matrix elements of $A$  
are obtained by expanding each
perturbed orbital in the basis of unperturbed orbitals. It can also be 
written in terms of the unperturbed and perturbed energy levels \cite{tanabe,hentschel2}.

According to the Friedel sum rule, the change in phase upon scattering at 
the local potential is related to the energy shift between the unperturbed 
and perturbed energy levels \cite{matveev}.
We therefore introduce the local phase shift for the level $i$ (with $i=M$ at the Fermi energy) as a measure of the  
rearrangement of levels
due to the perturbation as 
$$%\mbox{phase-shift}[M]
\phi_i=\pi \,\frac{\lambda_i\,-\epsilon_i}{\epsilon_{i}\,-\epsilon_{i-1}}.$$  
Here, we have assumed $V_c<0$. Similarly, for $V_c>0$ we find
$%\mbox{phase-shift}[M]
\phi_i=\pi \,(\lambda_i\,-\epsilon_i)/(\epsilon_{i+1}\,-\epsilon_i).$

%%%%%%%%%%%%%%%%%%%%%%%%%%%%%%%%%%%%%%%%%%%%%%%%%%%%%%%%%%%%%%%%%%%%%
\section{results}\label{result}
%%%%%%%%%%%%%%%%%%%%%%%%%%%%%%%%%%%%%%%%%%%%%%%%%%%%%%%%%%%%%%%%%%%%%

\subsection{AOC in the presence of degeneracies: Deviations from the 
Anderson power law}
\label{subsec_devpowerlaw}
%---------------------------
\begin{figure}
\includegraphics[width=10cm]{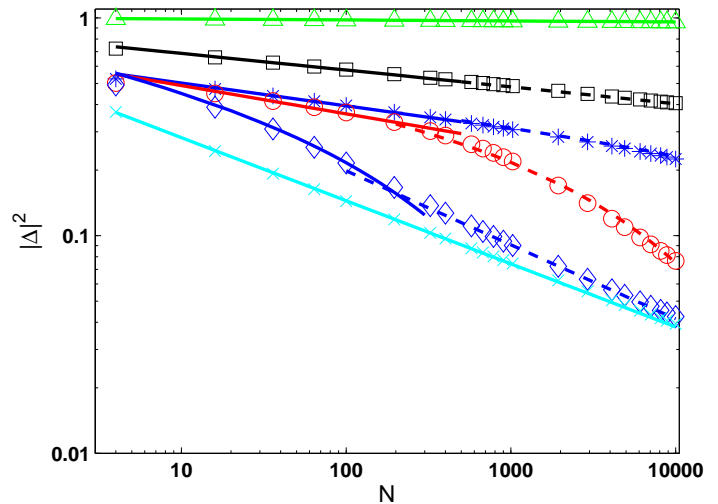}
\caption{(Color online) Many-body overlap $|\Delta|^2$ as a function of 
system size $N$ for six different strengths of perturbation. Points are 
obtained from the numerical evaluation of the Anderson overlap $|\Delta|^2$: the
different points $\triangle$,$\Box$, $\ast$, $\circ$, $\diamond$ and $\times$ are 
for  
$|V_c|/\omega_c=0.1, 1, 10, 100, 1000$ and  $10^{5}$ respectively. The solid and 
dashed curves represent the different functional form of $|\Delta|^2$ obtained from
analytics. The solid curves through $\triangle$, $\Box$, $\ast$, $\circ$ points are
of the form $N^{-0.5\delta^2/\omega_c^2}$ and well approximate the independent cluster 
regime (small $|V_c|/\omega$ and/or small $N$). Here $0 < \delta/\omega_c< 1$ denotes 
the shift in energy levels (phase shift) inside the last filled cluster. For $\triangle$, the perturbation being very weak 
(cf.~case (a) in the text), even for larger system size the power law beviour 
survives. For other three cases the power-law decay gets enhanced in presence of 
exponential decay $\exp(-\frac{3}{2}p^2\sqrt{N})$ in the large $N$ limit. Here 
$0<p<1$ is the measure of energy shift of the boundary levels in clusters. The 
$\diamond$ points show exponential decay even for small system size.
In this case, for large $N$ limit, overlap follows $N^{-\delta^2/\omega_c^2}$. 
Finally, a strong perturbation, case (b) in the text, yields the solid curve through 
$\times$ points, $N^{-\delta^2/\omega_c^2}$, and the Anderson power-law is recovered. 
Note that both $\delta$ and $p$ increase with $|V_c|$.  For 
$|V_c|/\omega_c=0.1, 1, 10, 100, 1000$ and  $10^{5}$ the power law exponents are 
$\delta^2/\omega_c^2=0.009 (\phi=0.095\pi), 0.18 (\phi=0.43\pi), 
0.19 (\phi=0.44\pi), 0.21 (\phi=0.46\pi), 0.0.26 (\phi=0.51\pi)$ and 
$0.26 (\phi=0.51\pi)$ respectively. For $|V_c|/\omega_c=1, 10, 100, 1000$ the 
coefficient of $\sqrt{N}$ in argument of the exponential functions are 
$\frac{3}{2}p^2=0.00005, 0.0008, 0.0112, 0.05$ respectively. Here we have 
chosen the magnetic control parameter $\omega_c=0.0001\omega_0$.}
\label{OvAnl}
\end{figure}
%--------------------------
First, we focus on the role of degeneracies on AOC. 
We therefore neglect the spatial dependence of the potential strength, i.e.,
we assume uniform amplitudes for all unperturbed orbitals throughout the PQD.
To this end we consider a PQD subject to a 
very weak magnetic field with frequency $\omega_c \ll \omega_0$ such that the
degeneracies in energy space are just lifted, but the shell structure is kept intact. 
We will refer to it as 
PQD in quasi-degenerate limit. 

For numerical calculations we have considered a PQD with a fixed number of
shells $S$, i.e., a total number $N=\frac{S\,(S+1)}{2}$ of levels with the 
lowest $M$ levels being occupied. Hereon, we will
refer to $M/N$ as `filling' of the PQD. 
In what follows, unless specified otherwise, we have chosen a PQD 
with a total number of shells $S=15$, i.e.,  total number of levels 
$N=120$.  All energies are measured in units of $\o_0$ ($\hbar=1$).

In Fig.~\ref{OvAnl} we show the behaviour of overlap $|\Delta|^2$ as a function of 
system size $N$ for six different $|V_c|$ keeping $\o_c$ and $\o_0$ fixed. 
We consider half-filled systems and $N$ such that the Fermi cluster is half-filled
 as well. In contrast to Anderson's result 
of a power-law decay of $|\Delta|^2$ as a function of $N$ for the bulk system, 
we find three different regimes in the presence of degeneracies (quasi-degenerate limit)
in the quantum dot. Depending on the strength of $|V_c|$, we observe (Fig.~\ref{OvAnl}):

(a) $|V_c|$ very small -- single cluster regime: $|V_c|$ is so small that the separation 
between the clusters remains large even after the perturbation is applied. The many-body 
overlap is dominated by the rearrangement of levels in the last filled cluster 
(i.e., the shell at the Fermi energy, the {\it Fermi cluster} in short). Thus we
can approximate the response of the whole system by that of the Fermi cluster,
which itself acts like a harmonic oscillator. 
If $s_M$ is the size of this cluster, the 
Anderson overlap is known \cite{tanabe, hentschel2} to be  
$|\Delta|^2 \sim s_M^{-(\phi/\pi)^2}$ where $\phi$ is the phase shift at the Fermi 
energy. It is easy to see that for a half-filled 
parabolic dot with in total 
$N$ levels, a half-filled Fermi cluster yields $s_M=\sqrt N$. Thus $|\Delta|^2 \sim N^{-0.5\,(\phi/\pi)^2}$. 
See the topmost two curves for small $|V_c|/\omega_c$ in Fig.~\ref{OvAnl}.

(b) $|V_c|$ large -- whole system responds: $|V_c|$ is so large that the shells mix and  
the whole system participates in the response. For large $|V_c|$, the inter-cluster
separation decreases to become comparable with intra-cluster level-separation. Then
the energy levels of all shells participate in AOC response.
In this regime we find the well-known result for the power-law decrease of 
the Anderson overlap, i.e.,  
$|\Delta|^2 \sim N^{(-\phi/\pi)^2}$. 

(c) $|V_c|$ intermediate -- transition regime: In this regime the results 
will depend crucially on the system size $N$ in addition to $|V_c|$. The gap 
between adjacent clusters decreases for bigger clusters, triggering the 
response of the whole system rather than a single shell. 
In addition, neighbouring clusters come closer to each other with increasing 
perturbation strength. This generates essentially 
the following sub-regimes in dependence on $N$ (assuming $|V_c|$ intermediate 
and constant):

(c-i) Small $N$: The separation between clusters is relatively large and one 
can again approximate the behaviour of the many body overlap of the whole 
system with that of the last partially (half-) filled cluster; thereby
getting back to the result for independent clusters as in the case (a) for 
small $|V_c|$: 
$|\Delta|^2 \sim N^{-0.5(\phi/\pi)^2}$. 

(c-ii) (Very) large $N$: For sufficiently large $N$, 
the clusters are close enough to each other such that all of them
contribute in the many-body overlap and give rise to the Anderson power law 
in this regime, similarly to case (b). In Fig.~\ref{OvAnl}, the many-body overlap  
for $|V_c|/\omega_c=1000$ follows 
$N^{-0.5(\phi/\pi)^2}\exp(-3p^2\sqrt{N}/2)$ in the small $N$ regime and decreases 
as $|\Delta|^2 \sim N^{-(\phi/\pi)^2}$ with $N$ for larger systems. Note that the 
local phase-shift  
$\phi=\pi\,\frac{\epsilon_M-\lambda_M}{\epsilon_M-\epsilon_{M-1}}$
at the Fermi energy is same for both $|V_c|/\omega_c=1000$  and 
$|V_c|/\omega_c=10^{5}$, although the overlap exhibits a very different 
power-law behaviour. The local phase-shift does not contain any information 
about the participating clusters. As a result, phase-shift becomes insufficient to fully 
describe the many-body overlap in PQDs. This, again, contrasts the behaviour 
known in bulk (metallic) systems 
(see \cite{metal}) where the phase shift at the Fermi energy alone 
is known to determine the overlap.  

(c-iii) Intermediate $N$: The overall behaviour is more complicated and characterised 
by the transition between the cases (c-i) and (c-ii). To 
gain a deeper, quantitative understanding, we have performed an explicit analytic 
calculation assuming that the last occupied shell is 
half filled and that this cluster and its two neighbouring clusters 
(one filled and the other empty) 
determine the behaviour of the many body overlap. This calculation is outlined in 
Appendix \ref{app} and yields the result
$|\Delta|^2 \sim N^{-0.5(\phi/\pi)^2}\,\exp(-3p^2\,\sqrt{N}/2)$. The solid curve 
through $\diamond$-points in Fig.~\ref{OvAnl} is provided by this expression. It well 
reproduces the overlaps in the transition regime, before the Anderson power law takes 
over again in large $N$ regime. 

To summarise the effect of the shell structure on the Anderson overlap for 
intermediate $|V_c|$ (cf.~the third, fourth curves from top in Fig.~\ref{OvAnl}),
 we expect the small $N$ regime with $|\Delta|^2 \sim N^{-0.5(\phi/\pi)^2}$
to cross over to an regime with $|\Delta|^2 \sim N^{-0.5(\phi/\pi)^2}\,\exp(-3p^2\,\sqrt{N}/2)$ for intermediate to large $N$. Increasing $|V_c|$ further, the clusters
start to mix even for small system size. In Fig.~\ref{OvAnl}, for 
$|V_c|/\omega_c=1000$, the many-body overlap $|\Delta|^2 \sim N^{-0.5(\phi/\pi)^2}\,\exp(-3p^2\,\sqrt{N}/2)$
in the small $N$ regime.  In the limit of very large $N$ the Anderson power law 
$|\Delta|^2 \sim N^{-(\phi/\pi)^2}$ is recovered. Note that for intermediate
strength of perturbation decay rate of many-body overlap with system size
could be faster than in case of stronger perturbation.  
This change in power law is due to the number of participating clusters in 
the many-body overlap, note that the phase shift $\phi$ at the Fermi energy 
itself does not change. This behaviour marks a clear deviation from the 
well-known Anderson result \cite{anderson} for bulk systems that was also 
found to hold for other (chaotic) mesoscopic systems \cite{gefen1,gefen2,
hentschel1,hentschel2}. It is a consequence of the shell structure, i.e., the 
degeneracy of energy levels. Since this can be easily realized in the mesoscopic 
regime by using the well-established semiconductor-heterostructure parabolic 
quantum dots, this interesting behaviour becomes accessible to experiments.

\subsection{Shell effects in AOC} 

%-------------------------------
\begin{figure}
\includegraphics[width=9cm]{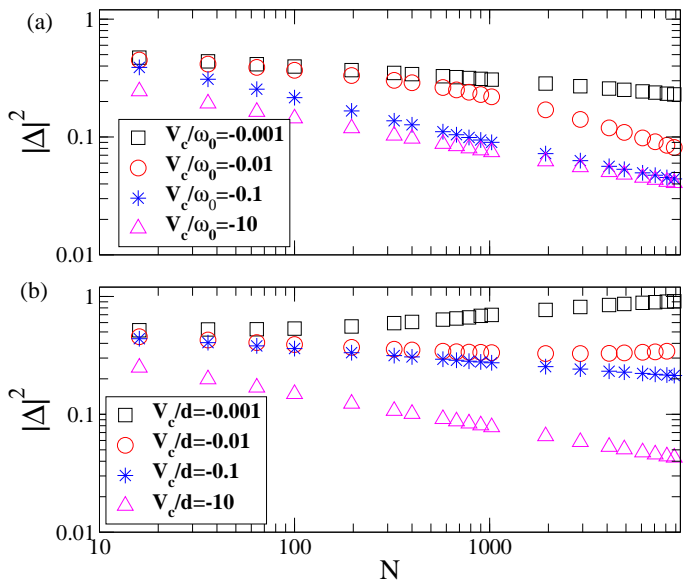}
\caption{(Color online) (a) Manybody overlap $|\Delta|^2$ as
a function of system size $N$ for $4$ different strength of perturbation
scaled with oscillator's frequency $\o_0$ is compared with (b) $|\Delta|^2$ 
as a function of system size $N$ for four different strengths 
of perturbation scaled with average level spacing $d$.} 
\label{compVc} 
\end{figure}
%--------------------------------

%-------------------------
\vskip 0.5cm
\begin{figure}
\includegraphics[width=8cm]{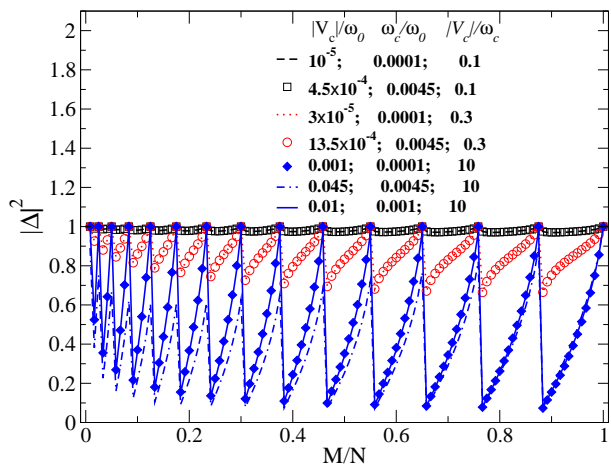}
\caption{(Color online) Many-body overlap  $|\Delta|^2$ for a PQD is shown
 as a function of filling $M/N$.  In the regime of weak magnetic 
 field ($\omega_c < 0.01\omega_0$) and weak perturbation $V_c\le 0.01\omega_0$, the magnitude of overlap remains same for the same ratio of $|V_c|/\omega_c$. }
 \label{scale}
\end{figure}
%---------------------------
\begin{figure}
\includegraphics[width=8cm]{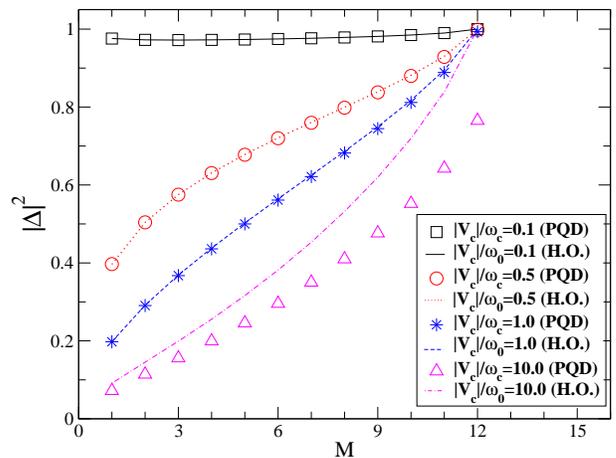}
\caption{(Color online) Many-body-overlap $|\Delta|^2$ as function of 
number of electrons $M$. The $12$-th cluster of a PQD with
$15$ shells (symbols) is compared with $|\Delta|^2$ of a $1D$ harmonic 
oscillator (HO) having $12$ energy levels with level spacing $\o_c$ (lines). 
The PQD is characterised by $\omega_c=0.0045\o_0$, and $M$ refers to the 
number of electrons in the 12th shell. Deviations between the two curves 
occur away from the small $V_c/\o_c$ limit.
}
\label{comp1dpqd}
\end{figure}
%---------------------------

Another illustration that highlights the role of shells and the $|V_c|$-dependent 
inter-shell spacing is shown in Fig.~\ref{compVc}. We consider $|\Delta|^2(N)$ 
for different $|V_c|$ that are scaled with respect to the inter-shell spacing measure 
$\o_0$ in Fig.~\ref{compVc}(a) and with respect to the mean level spacing $d$ 
(defined using only the levels in occupied clusters, $d=2\Omega/s_M$ in the large 
dot limit) in Fig.~\ref{compVc}(b). The motivation behind is a scaling of the 
Anderson overlap with $|V_c|/d$ that was reported in Refs.~\cite{hentschel1,hentschel2}. 
For intermediate 
perturbation strength (squares in Fig.~\ref{compVc}),  
$|\Delta|^2$ decreases with $N$ in (a) but increases with $N$ in panel 
(b) in the large cluster limit. Since $d$ is inversely proportional 
to $\sqrt{N}$, for larger clusters to keep the ratio $|V_C|/d$ constant 
one needs to apply a weaker perturbation. This generates a larger many-body 
overlap with increasing system size. Consequently, the scaling of physical 
quantities with $|V_c|/d$ that was found to be characteristic for other 
(chaotic) mesoscopic systems, can be completely 
overtaken by the shell structures in PQDs. 

\begin{figure*}
\includegraphics[width=14cm]{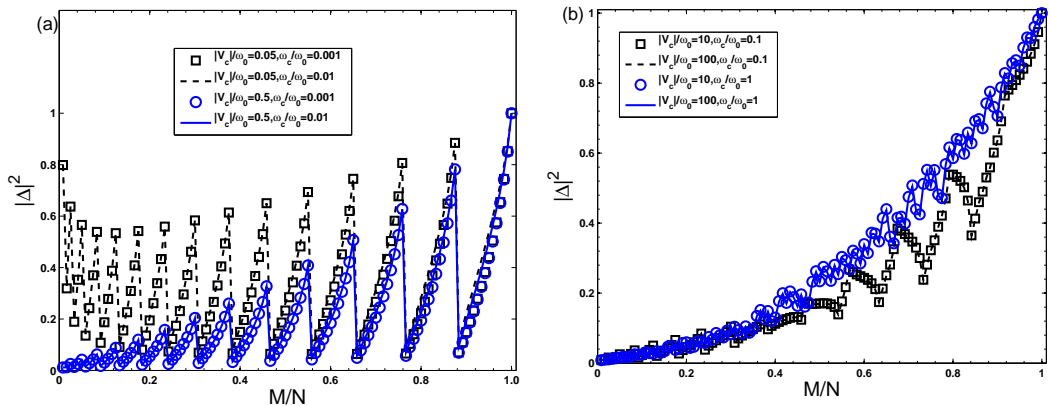}
\caption{(Color online) Many-body overlap  $|\Delta|^2$ for a PQD shown
 as a function of filling $M/N$. The results are obtained for (a) moderate 
perturbation $|V_c|\ge 0.01\omega_0$ in the quasi-degenerate limit and 
(b) for strong magnetic 
field ($\omega_c \ge 0.1\omega_0$) and strong perturbation $|V_c| > 5\omega_0$.
 In (a) the magnitude of $|V_c|$, determines the value of the overlap at all 
filling. In (b), behaviour of the overlap is controlled by $\o_c$.}
 \label{BOvlrg}
\end{figure*}
%---------------------------
\begin{figure}
\includegraphics[width=8cm]{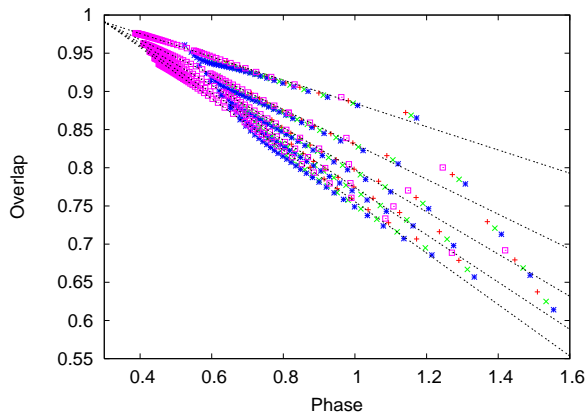}
\caption{(Color online) The many-body overlap as a function of phase shift
for three different PQDs as well as for a number of $1D$ harmonic oscillators 
(HO). The different symbols are: PQD with 
$15$ ($+$), $25$ ($\times$), $35$ ($*$) clusters and HOs with 
$1$ up to $35$ levels ($\Box$). The dashed lines are guide to the eye. 
From top to bottom, the five different curves correspond to 
clusters of a PQD and HOs with 
$1,2\cdots 5$ unoccupied levels, respectively. 
For the PQD, the magnetic control is chosen as 
$\o_c=0.0001\o_0$ the frequency of harmonic oscillator is 
$\o_c$. All $4$ systems are perturbed by an attractive perturbation of
strength $|V_c|=0.0001\o_0$.
}
\label{OvPh}
\end{figure}
%-----------------------------
The possibility to precisely control the number of electrons on quantum dots 
through Coulomb blockade motivates us to study the many-body overlap 
$|\Delta|^2$ as a function of filling $M/N$, where $M$ is the number of filled
energy levels and $N$ is the total number of energy levels considered. 
The results obtained in the quasi-degenerate limit 
are shown in Fig.~\ref{scale}. 
We consider very weak perturbations ($|V_c|\le 0.01\o_0$) and tune 
$|V_c|$ and the magnetic control field $\omega_c$. 
From Fig.~\ref{scale}, 
one can see that changing $|V_c|$ and $\o_c$ keeping the ratio 
$|V_c|/\o_c$ constant generates the same magnitude 
of overlap at a given filling $M/N$, i.e., a scaling behaviour of $|\Delta|^2$ holds for 
small $|V_c|/\o_c$. The shell structure with 15 distinct clusters is easily 
visible in  Fig.~\ref{scale}. 
The overlap reaches $1$ at the completion of each cluster;  as the 
adjacent clusters - for the small $|V_c|$ considered here - are widely 
separated in energy and do not contribute. Note that the overlap drops 
to a {\it small} value whenever a new shell is opened, even for the 
{\it weak} $|V_c|$ considered here. The reason is the larger 
size of the available phase space for shells with small filling. The drop in 
the many-body overlap at the opening of a new shell is a phase-space factor that - for sufficiently 
small perturbations - is given by the inverse of the level degeneracy. Details will be discussed 
elsewhere \cite{georg}.

Interestingly, for weak perturbations ($|V_c|/\omega_0 \le 0.01$, 
the quasi-degenerate limit) the many-body overlap $|\Delta|^2$ of each 
cluster behaves independently and like an $1D$ harmonic oscillator (H.O.)
of natural frequency $\o_c$ 
and the same number of levels as in the cluster in consideration 
subject to the same perturbation strength $|V_c|$, see Fig.~\ref{comp1dpqd}. 
For the $1D$ 
harmonic oscillator, the natural frequency gives the measure of level-spacing
 whereas for PQD the same role is played by the cyclotron frequency $\o_c$. 
 In this case, applying a weak magnetic field,
the adjacent clusters of a PQD remain well separated in energy. The
weak perturbation $|V_c|$ shifts these clusters only slightly. As a result
each cluster behaves independently or, in other words, the many-body overlap is 
determined by the last (partially) filled cluster alone.

In Figs.~\ref{scale} and \ref{comp1dpqd} we have focused on the case of weak 
perturbations and a small (control) magnetic field. We now increase the perturbation 
strength $|V_c|$ to values that appear to be more realistic, e.g., in the context of 
photoemission- or photoabsorption-induced Fermi edge singularities (see Fig.~\ref{BOvlrg}).
The chosen $\omega_c$ scale has been increased a little such that it has a common 
$\omega_c$ regime with Fig.~\ref{scale}. 
In this intermediate regime of perturbation, the shell structure response is 
somewhat modified, for example, now the overlap remains smaller than one even 
at the complete filling of a shell - a direct consequence of AOC. 
Note in particular that the overlap does not scale with 
$|V_c|/\omega_c$ anymore, see Fig.~\ref{BOvlrg}(a). Rather, it is now 
determined by $|V_c|$ alone (remember that we set $\o_0$ constant), 
and is {\it not} affected by changes of $\o_c$. 

Next we turn on a strong magnetic field such that clustering of energy 
space is broken partially or completely (see Fig.~\ref{BOvlrg}(b)), and 
also increase $|V_c|$ further. With increasing $\o_c$ the levels in an unperturbed 
cluster start to spread apart from each other and eventually mix with levels 
from the neighbouring clusters. Accordingly, the shell structure in 
$|\Delta|^2(N)$ is now (almost) lost, and the overlap shows an approximate 
monotonic increase with filling. 
In this regime, the behaviour of $|\Delta|^2$ as a function of filling is
solely determined by the magnetic field strength (characterised by cyclotron
frequency $\o_c$). However a fine tuning of $|\Delta|^2$ is possible by 
changing $|V_c|$.

It would be worthy to mention that again at around $\o_c=0.7\o_0$ which 
corresponds to a new clustering of unperturbed levels  (see lower panel 
of Fig.~\ref{engsp} and the discussion there) all three distinct regimes 
of $|\Delta|^2$ as a function of filling can be observed over a wider span 
in $|V_c|$-$\o_c$ parameter space. 

We close this subsection by a discussion of shell effects and the scaling 
behaviour observed in Fig.~\ref{scale} in terms of phase shifts. It is well 
known that for metals,  the AOC (and FES) response is fully 
characterised by the phase shift at the Fermi energy, through the Anderson 
power law \cite{remarkFES}. 
Here we study the many-body overlap as a function of phase shift at the Fermi
energy in the presence of 
shell structures (see Fig.~\ref{OvPh}).
We have collected the many-body overlap from partially filled clusters of 
three different systems with $15, 25$ and $35$ shells, respectively, 
characterised by different symbols in Fig.~\ref{OvPh} subject to a (very) 
weak perturbation (similar to Fig.~\ref{scale}). We have considered five 
different fillings for each shell, namely each shell contains $1,2,\ldots 5$ 
unoccupied levels.
From Fig.~\ref{OvPh}, one can easily see that five different curves 
emerge from the collected data, each of them corresponds to
a fixed number of unoccupied levels. As expected from Fig.~\ref{scale}, the 
overlap is larger for `almost filled' shells and it decreases with the 
increasing number of empty levels. In each curve, the overlap increases with 
increasing size of clusters or filling of clusters. The dashed straight 
lines are a guide to the eye. This observed linearity is a manifestation 
of scaling behaviour of $|\Delta|^2$ for constant $|V_c|/\o_c$ of a PQD 
subject to weak perturbation in the quasi-degenerate limit. For comparison, 
we have calculated overlap and phase-shift at Fermi energy for harmonic 
oscillators with 35 equidistant energy levels, among which $1,2,\cdots 5$ 
are empty. In this weak perturbation regime, as expected, the results 
(marked as boxes in Fig.~\ref{OvPh}) match with the five curves obtained 
for PQD. Note that the phase shift decreases as we go to higher clusters 
(i.e., higher fillings). Deviations from linearity occur as soon as the 
different clusters start to mix.
Note that for a PQD with $15$ shells ($N=120$) the mixing occurs at around
 $\o_c=0.07\o_0$, whereas for a PQD with $25$ shells ($N=325$) and $35$ shells
($N=630$) mixing starts at much lower $\o_c$. This is the reason that for the 
PQD with $15$ shells linearity holds for all the clusters present in the system but for larger systems linearity breaks for the higher (and therefore larger) 
clusters.

%%%%%%%%%%%%%%%%%%%%%%%%%%%%%%%%%%%%%%%%%%%%%%%%%%%%%%

\begin{figure}
\includegraphics[width=9cm]{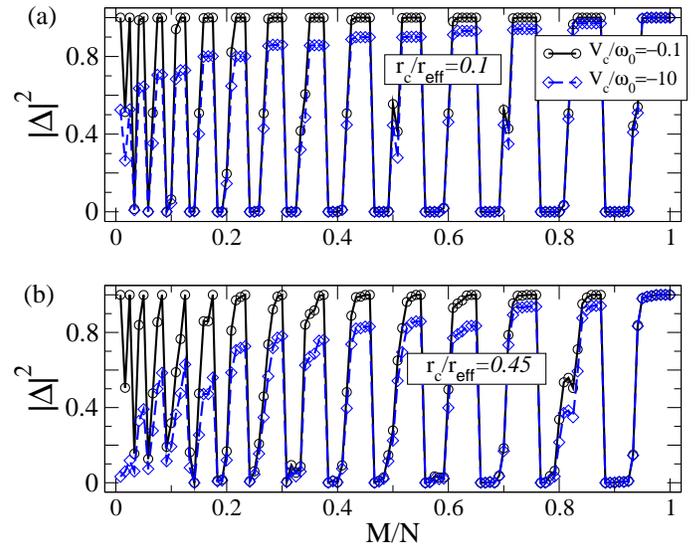}
\caption{(Color online) Many-body overlap $|\Delta|^2$ as a function of 
filling $M/N$ for a PQD with $15$ shells ($N=120$) under intermediate
($\circ$) and strong ($\diamond$) perturbation applied (a) very close
to the PQD's centre and (b) away from centre.
} 
\label{MesOvFill}
\end{figure}
%---------------------------
\begin{figure}
\includegraphics[width=8cm]{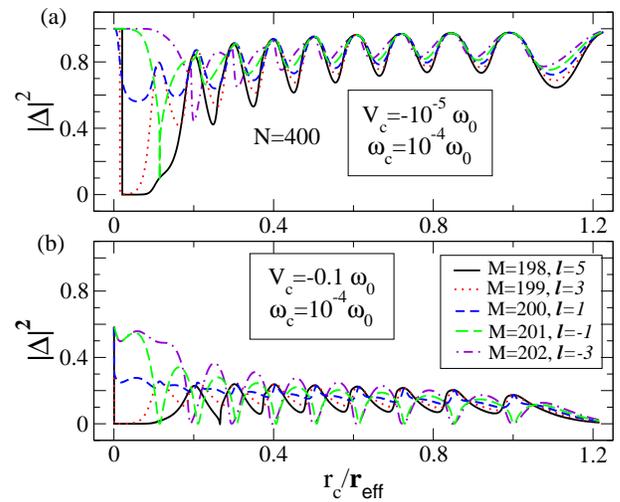}
\caption{Many-body overlap $|\Delta|^2$ as a function of the position $r_c$ of 
the perturbation %core hole $r_c$ 
for a PQD ($N=400$) around half-filling under (a) weak and (b) moderate 
attractive perturbation.
} 
\label{OvrcFL}
\end{figure}
%-------------------------

\subsection{AOC in the mesoscopic case}
%%%%%%%%%%%%%%%%%%%%%%%%%%%%%%%%%%%%%%%%%%%%%%%%%%%%%%

So far we have focused on the impact of the shell structure on  
AOC and assumed constant wave-function amplitudes 
(as in the metallic case) throughout the PQD. Now we turn to the truly 
mesoscopic case, i.e., the position dependence of the wave-function amplitude 
at the point of perturbation will be taken into account.

First we consider a PQD with $15$ shells (i.e., $N=120$) under 
intermediate and strong perturbation applied at two different positions. 
We calculate the many-body overlap $|\Delta|^2$ 
as a function of filling of the PQD, as before,
 in Figs.~\ref{scale} and \ref{BOvlrg}. The results are shown in 
Fig.~\ref{MesOvFill}. As before, here too, one can clearly identify 
the existence of the $15$ shells in the dot. 
However, unlike the previous case, within a shell the overlap $|\Delta|^2$ 
may now be non-monotonic function of filling
and often forms a plateau at the shell 
boundaries. This saturation of the overlap is an interesting feature 
especially from the point of view of experimental studies.

\begin{figure}
\includegraphics[width=6.9cm]{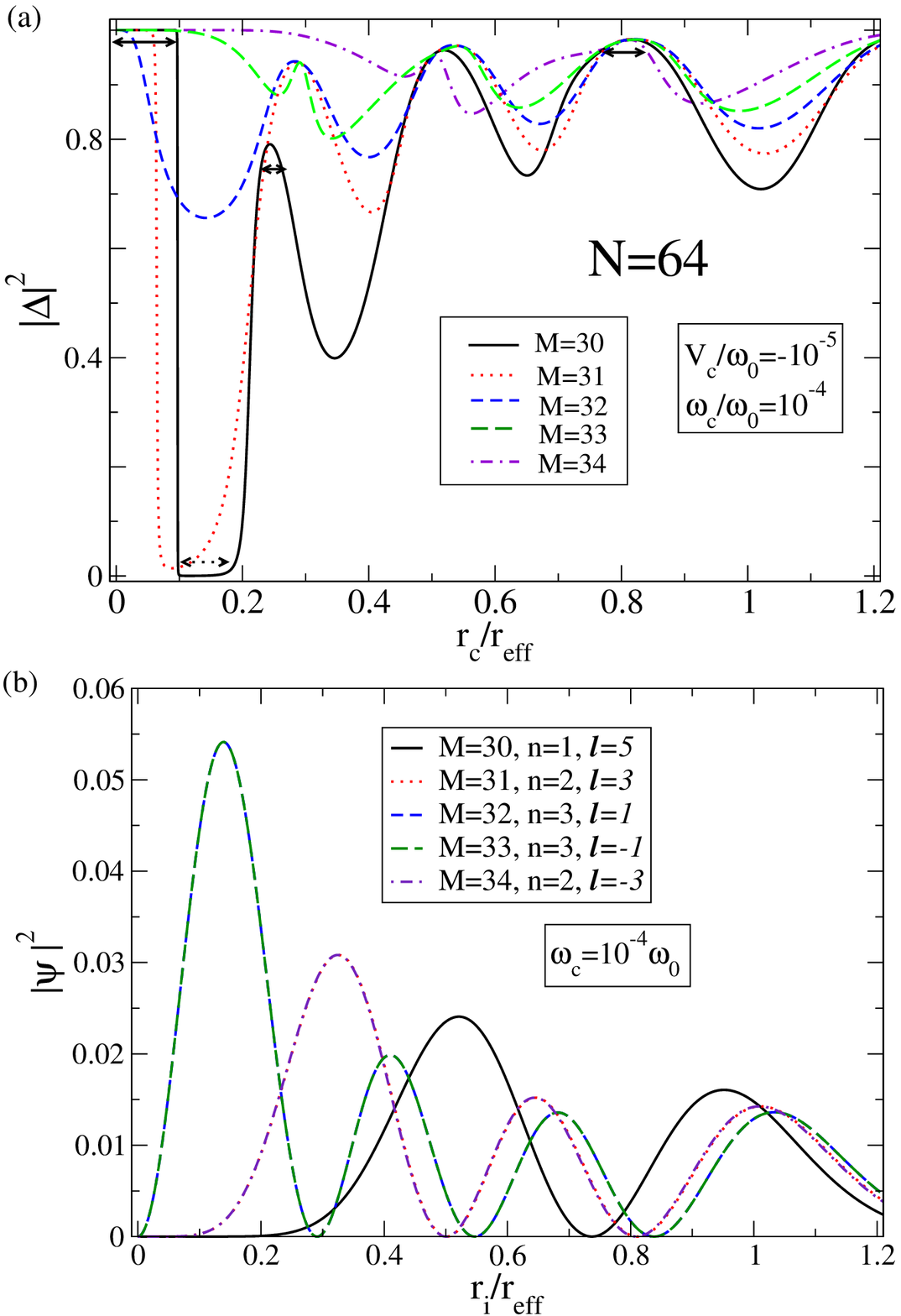}
\vskip 0.2cm
\includegraphics[width=7cm]{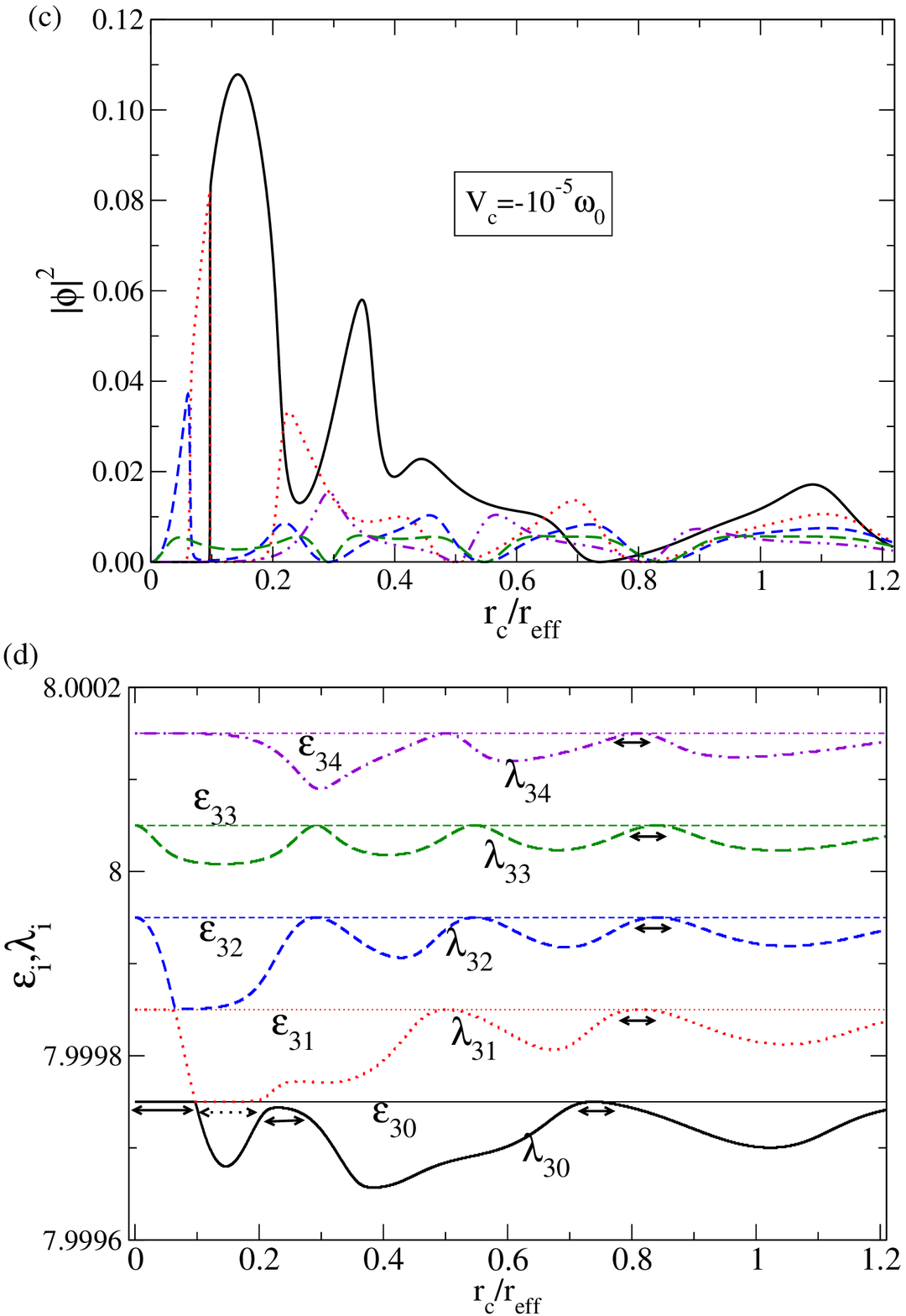}
\caption{(Color online) 
(a) Many-body overlap $|\Delta|^2$ as a function of 
$r_c$ for a PQD ($N=64$) around half filling ($M=32\pm 2$) subject
to a weak attractive perturbation.  
(b) Unperturbed 
orbitals as a function of radial distance $r_i/r_{\mbox{eff}}$ from 
the centre of the same PQD.
In (c) and (d) the amplitude of orbitals {\it at the position of the 
perturbation} and the corresponding perturbed energy levels are shown 
respectively. The straight lines in (d) mark the position of the unperturbed 
energies $\epsilon_i$. See text for details.
} 
\label{WvPuP}
\end{figure}
\begin{figure}
\includegraphics[width=7cm]{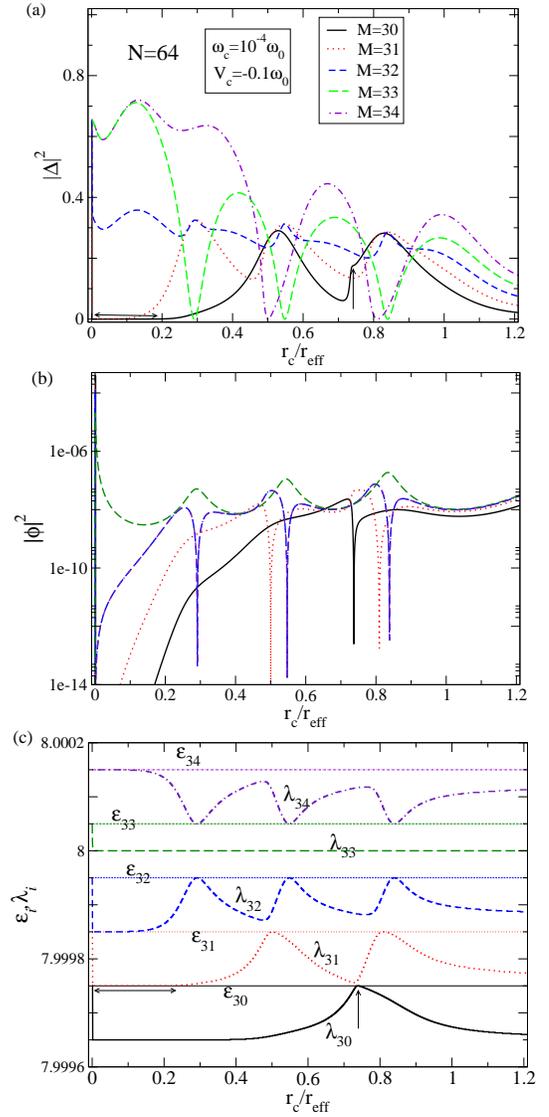}
\caption{(Color online) (a) Many-body overlap $|\Delta|^2$ as a function of 
$r_c$ for a PQD ($N=64$) around half filling ($M=32\pm 2$) subject
to an attractive perturbation of moderate strength. In (b) and (c),  
the amplitude of orbitals {\it at the position of the perturbation} and the 
corresponding perturbed energy levels are shown respectively. The horizontal 
straight lines in (c) mark the position of the unperturbed energies 
$\epsilon_i$.}  
\label{WvPuPS}
\end{figure}

Next we study how the Anderson overlap varies with position $r_c$ of the 
localised perturbation. To this 
end we consider PQDs with $N=64$ and 
$400$ around half-filling 
$M/N=1/2$ deep inside the quasi-degenerate limit and  apply weak and 
moderate perturbations. Again, $N$ is chosen such 
that for a half-filled system, the Fermi cluster is also half filled. 
For the smaller $N=64$ (bigger $N=400$) dot the last partially filled 
cluster is the $8$th ($20$th) cluster. We vary the
 position of the perturbation and calculate many-body overlap $|\Delta|^2$ 
for given $|V_c|$ and $\o_c$. 

The result is shown in Fig.~\ref{OvrcFL} for (a) a weak and (b) a moderately strong 
perturbation in the large dot ($N=400$). In both cases, the overlap varies with position $r_c$ and shows strong dependence on filling $M/N$. 
These oscillations depend on the structure of the wave functions, in 
particular on the angular momentum quantum number $l$, as the 
effective perturbation felt by the system depends on the wave function 
amplitude at $r_c$ around the Fermi energy, besides the strength 
of $|V_c|$. The value of $l$ determines, in particular, the behaviour 
near the center of the dot. We discuss it in more detail below. The overlap is, 
as expected, smaller for the larger perturbation for most $r_c$. 
Note, however, that unlike the bulk (or bulk-like) case, even within
 a cluster, the overlap is not a monotonic function of filling (or, $M$ for fixed $N$).
 
To understand the oscillatory behaviour of the overlap,
we have chosen a smaller dot with $N=64$ 
subjected to weak and intermediate perturbations in the quasi-degenerate limit, 
see Figs.~\ref{WvPuP} and \ref{WvPuPS} respectively. 
We consider again five fillings near the half filling.  
The corresponding highest filled unperturbed orbitals are shown in Fig.~\ref{WvPuP}(b). 
They all have nodes at the centre, as $l \neq 0$, with a wider dip for larger $l$. 
The value of the perturbed orbital, at the position of the perturbation, is shown in 
Figs.~\ref{WvPuP}(c) and \ref{WvPuPS}(b). It illustrates the strong effect of the 
perturbation on the orbital intensity, especially near the dot-centre, 
as well as the differences between weak and intermediate perturbations. 
The lowest panels, in both the figures, show the perturbed single particle 
energies, again as a function of $r_c$. 

The  Anderson overlap $|\D|^2$ in Figs.~\ref{WvPuP}(a) and \ref{WvPuPS}(a) 
can be understood by referring to Eq.~\ref{OvorgF}.
When $\lambda_{M+1}$ equals 
to $\epsilon_M$, many-body-overlap $|\Delta|^2$ goes to zero 
(indicated by dotted horizontal arrows near $r_c=0.15$ in 
Figs.~\ref{WvPuP}(a) and ~\ref{WvPuP}(d)).  
On the other hand, when $\lambda_{M}$ equals to $\epsilon_M$, 
the many-body overlap goes to $1$.
Similarly, when the unperturbed last filled level hardly changes after 
perturbation, $|\Delta|^2$ shows a peak, 
marked by solid horizontal arrows in Fig.~\ref{WvPuP}. Thus for weak perturbations,
 the filled and empty levels close to Fermi energy largely determine the many-body overlap.
In the case of stronger perturbations (Fig.~\ref{WvPuPS})  
the values for the overlap are smaller in general. 
When the last filled energy level coincides with the next 
higher perturbed level, $|\Delta|^2$ goes to zero as before (marked by the solid horizontal arrows). 
However, in general the behaviour is more involved, note in particular the larger spread in the overlaps 
at a given $r_c$.  Unlike the weak perturbation case, 
not only few levels close to Fermi-level are 
dominating the many-body overlap. Rather, more and more levels play a significant role in 
determining $|\Delta|^2$. In other words, the AOC many-body effect becomes more
prominent with stronger perturbation. 
Note that the large fluctuations in 
$|\Delta|^2$ as well as the spikes in $|\phi|^2$ are related to the occurrence of avoided level crossings at the Fermi 
energy, one example is marked by the solid vertical arrow in Fig.\ref{WvPuPS}(c).

\begin{figure}
\includegraphics[width=8cm]{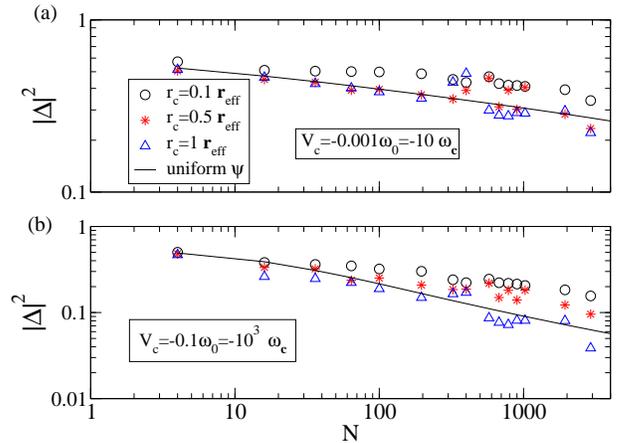}
\caption{(Color online) (a) Manybody overlap as a function of $N$. We consider 
half-filled clusters and half-filled systems. (b) For comparison, the overlap 
obtained for uniform amplitude is also shown.
}
\label{MesOvVsN}
\end{figure}

We now investigate the system size dependence of the many-body-overlap 
$|\Delta|^2(N)$ for three different positions $r_c$ of the perturbing 
potential. We consider half-filled systems 
with half-filled Fermi clusters, in order to avoid 
overshadowing
effects that might have come from differences in shell-fillings.
Two sets of results, obtained for two different
$|V_c|$, are shown in Fig.~\ref{MesOvVsN}. When the perturbing position is chosen 
close to the centre of 
the dot where most of the wave-functions (all with non-zero $l$) have zero amplitude, 
the overlap is indeed somewhat larger than in the uniform amplitude case (see the open circles in 
Fig.~\ref{MesOvVsN}). 
Applying the perturbation away from the dot centre, i.e., away from the 
centrifugal barrier (triangles and stars), makes the presence of mesoscopic fluctuations 
evident: $|\Delta|^2$ shows a strong dependence on $r_c$ at each $N$. 
Consequently, there are 
deviations from the uniform amplitude case keeping the qualitative nature unchanged.
The deviations would be averaged out after averaging over a large number of realizations.
 Accordingly, the deviations from the Anderson power law exponent, discussed in Section 
\ref{subsec_devpowerlaw}, apply now to the averaged many-body overlap.

\begin{figure*}
\includegraphics[height=6cm,width=13cm]{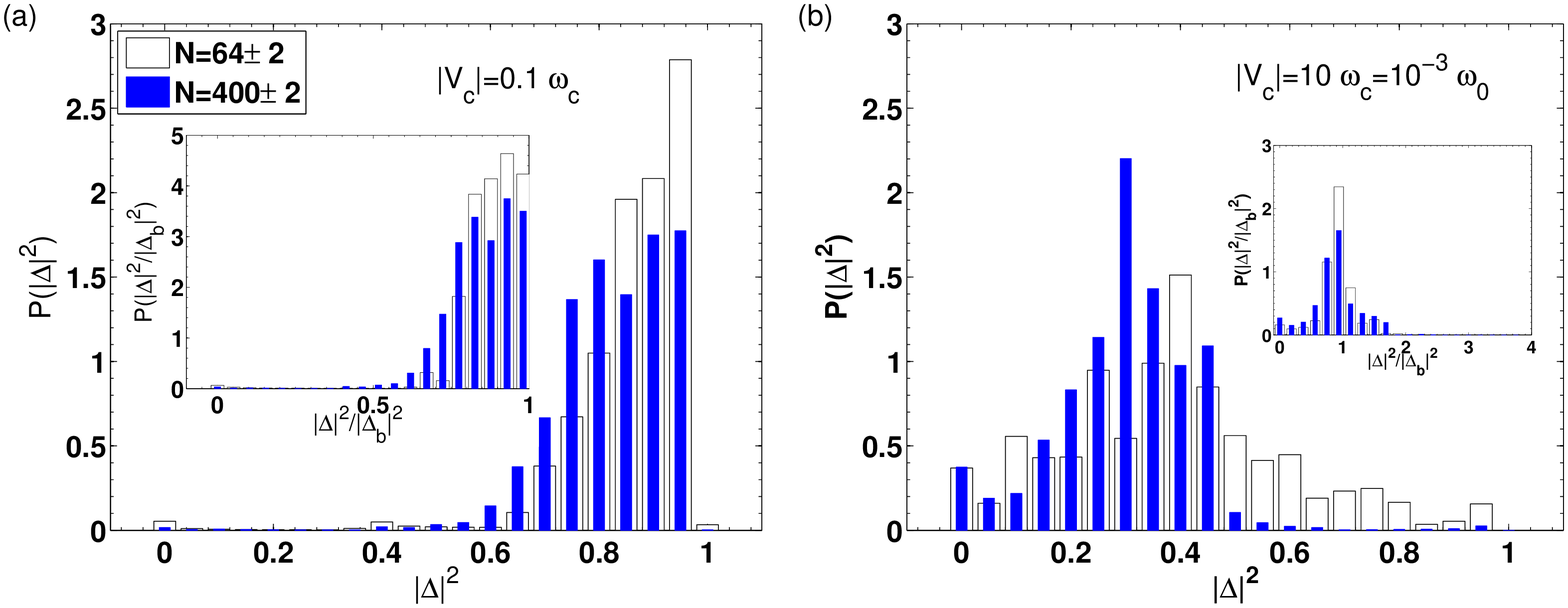}
\includegraphics[height=6cm,width=13cm]{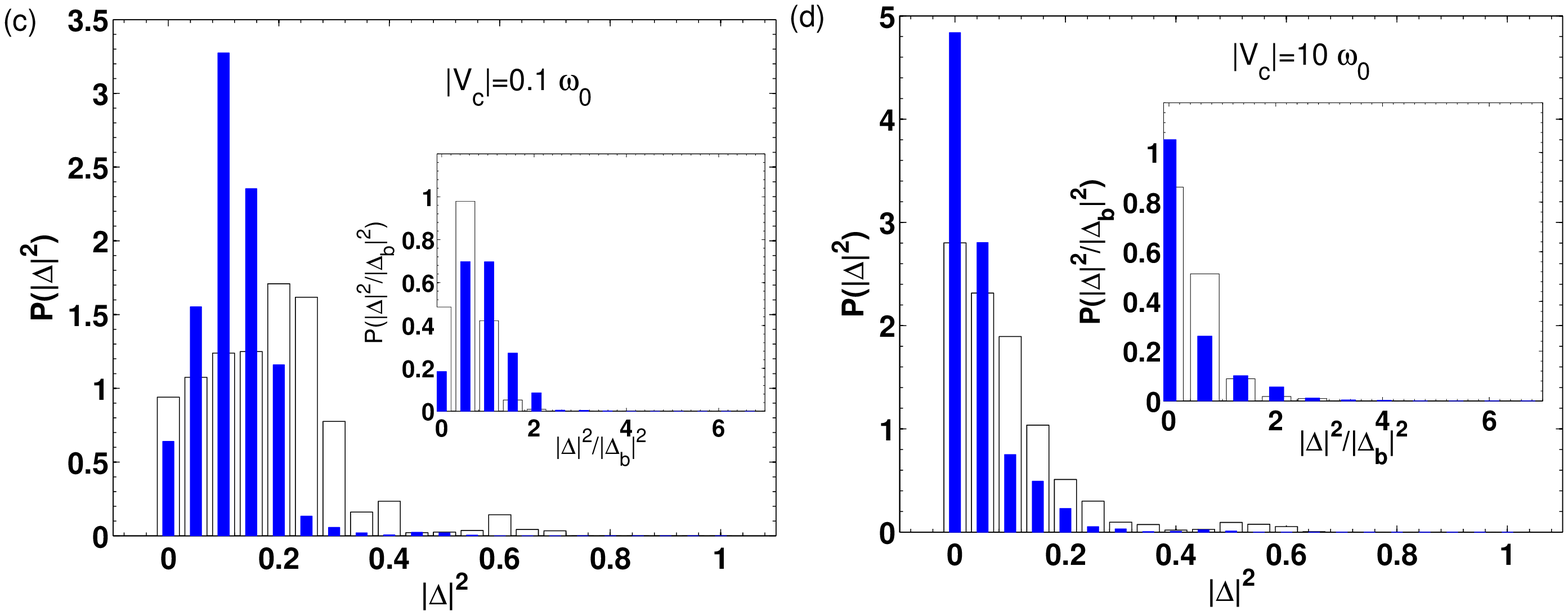}
\caption{(Color online) The distribution of many-body overlaps for 
two different PQD sizes (open and solid bars) and  
half-filled systems under (a) weak, (b) moderate, (c) strong, 
and (d) very strong attractive perturbation. 
To obtain the distribution, data were collected over $5$ different $M$ 
close to half-filling as well as  $5$ different values of $N$ such that 
$N\in [2M-2,2M+2]$. $5000$ different $r_c$ were chosen for 
each pair of $M, N$. We set $\omega_c=0.0001\omega_0$.
In the inset of each panel, the corresponding distribution of the overlap 
scaled by $\Delta_b$ is shown. Here $|\Delta_b|^2$ is the many-body overlap 
with uniform wave-amplitude in PQD.}
\label{stat1}
\vskip 0.5cm
\end{figure*}

Now we focus on the mesoscopic fluctuations and discuss 
the probability distribution of overlaps $|\Delta|^2$ for 
two systems around half filling (Fig.~\ref{stat1}). We have chosen 
four different strengths $|V_c|$ of attractive perturbation, 
its position $r_c$ is assumed to be uniformly spread over the PQD and collected over $5000$ 
different values to estimate the overlap distribution. 
Because of the circular symmetry of the PQD $|\D|^2$ is independent of the
angular positioning of the perturbation and depends only on $r_c$.
For the weak perturbation case, as expected, the distribution 
$P(|\Delta|^2)$ has a large peak near $|\Delta|^2=1$ (see Fig.~\ref{stat1}(a)). 
Since, even for weak perturbation the overlap can be very small when a cluster is less
than half filled, 
one can find a very small but finite probability 
for very small overlaps. Increasing the strength of the 
perturbation, the overlap distribution becomes even wider before reducing in width towards 
very strong perturbations. Getting a larger overlap becomes less and less
probable and the peak is moved to smaller values (Fig.~\ref{stat1}(b), (c) and (d)). 

It is interesting to compare the distributions for $N=64$ and $N=400$, e.g., 
in Fig.~\ref{stat1}(b). The smaller system possesses the larger overlaps, as 
expected. If this is taken into account by scaling all results by the overlap 
$\Delta_b$ of the corresponding PQD with uniform amplitude (see insets), the 
two probability distributions coincide much better, though not as good as in 
Refs.~\cite{hentschel1,hentschel2}. 
 
\begin{figure}
\includegraphics[width=8cm]{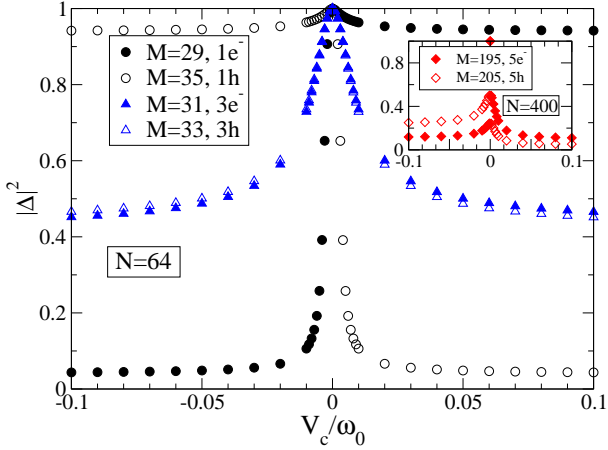}
\caption{(Color online) The many-body overlap as a function of perturbing 
potential strength $V_c$ is shown for a few selected $M$ chosen from $8$-th 
cluster of a PQD having $64$ levels in total. Here, the symbol e${}^-$ stands 
for extra (from nearest closed cluster) electrons, h for extra holes. 
We set $\omega_c=0.0001\,\omega_0$ and 
$r_c/r_{\mbox{eff}}=0.5$. In the inset, a pair of curves representing many-body overlap
as a function of perturbing potential strength $V_c$ are shown for a PQD with 
$N=400$ levels and $M$ belong to $20$-th cluster.}
\label{OvVc}
\end{figure}

Finally, we address the particle-hole symmetry of the Anderson overlap
(see Fig.~\ref{OvVc}). We calculate the many-body overlap for few selected 
fillings in the $8$-th cluster of a PQD 
with $N=64$ that are related by particle-hole symmetry. 
For not too large $|V_c|$, 
an electron in the presence of an attractive perturbation and a hole in
the presence of a repulsive perturbation represent the 
 same thing physically. 
From Fig.~\ref{OvVc}, one can see that
the overlap for a fixed number excess electrons (number of extra electrons 
from closest completely filled cluster) is mirror symmetric about 
$V_c=0$ to the overlap for a PQD with same number of holes (deficit in number 
of electrons to form a completely filled cluster). However, for stronger
 perturbations, when different clusters start to mix, the particle-hole 
symmetry does not hold. For bigger systems, as the mixing starts for weaker 
perturbation (as we have seen in Fig.~\ref{OvAnl}) the asymmetry in many-body 
overlap appears for much smaller strength of perturbation.
 With increasing $V_c$ (positive or negative), $|\D|^2$ 
decreases. However, the reduction, beyond the symmetric regime of small $V_c$, 
is much steeper in the case of repulsive $V_c$ than for attractive $V_c$.  

%%%%%%%%%%%%%%%%%%%%%%%%%%%%%%%%%%%%%%%%%%%

\section{Conclusion}\label{concl}
We have studied the Anderson orthogonality catastrophe (AOC) for experimentally accessible parabolic quantum dots. 
Our results show that in addition to strength and position of 
the perturbation, level degeneracy and the rearrangement of shell structures
are important in determining the behaviour of the many-body overlap.

As a reference system, we started with a simplified PQD with 
uniform amplitudes. For such a system, shell effects in the Anderson overlap 
are strong, especially in the quasi-degenerate limit. This shell effect 
survives in the mesoscopic case as well, even in presence of mesoscopic 
fluctuations. The 
rearrangement of energy levels 
between the shells 
leads to deviations from the well-known Anderson power law of the overlap. 
It is especially prominent for intermediate perturbation strengths when the power 
law behaviour of independent cluster approximation in small $N$ regime gets 
modified by an exponential decay with system size in the large $N$ regime 
(Fig.~\ref{OvAnl}).

In the mesoscopic situation with position dependent wave-function amplitudes, 
the position of the perturbation strongly influences the Anderson overlap. 
We obtained a broad distribution of overlaps for various positioning of the 
localized perturbations. This distribution scales with the overlap $|\Delta_b|^2$ obtained for a system with uniform amplitudes. 

Our results underline the richness of samples in the mesoscopic regime with 
respect to many-body physics. In the present study, it is in particular the 
degeneracy of levels in parabolic quantum dots that makes AOC
 an interesting topic with new features that go 
well beyond the well understood bulk metallic case. 

\appendix
%\section{Different participating clusters and power-law behaviour of overlap}
\section{Power-law behaviour of the overlap in dpendence on perturbation strength and participating clusters}
\label{app}
For a contact-type or rank-one perturbation, the many-body overlap can be 
expressed in terms of all single
particle energy states~\cite{tanabe} as
\begin{equation}
	|\Delta|^2 = \prod_{i=1}^M\,\prod_{j=M+1}^N\,\frac{(\lambda_j-\epsilon_i)\,(\epsilon_j-\lambda_i)}
	{(\lambda_j-\lambda_i)\,(\epsilon_j-\epsilon_i)}\,.\label{OvorgF}
\end{equation}
Depending on the perturbation strength, the lowest level $\lambda_0$ of a chaotic system 
gets shifted by a larger amount (`` formation of a bound state'') than the other 
levels that are bound by their neighbors. 
For PQD, this scenario applies to the lowest level of each cluster that can be shifted  more compared to all other levels of the cluster in application
of the same perturbation. For example, the lowest level of Fermi cluster ($s_M$) can have a shift of order $p\,D_{s_{M-1}}$
with $0<p<1$ and $D_{s_{M-1}}=\omega_0-s_M\,\omega_c+3\omega_c/2$, the energy gap with the neighbouring
filled cluster ($s_{M-1}$).  All other levels of the cluster ($s_M$) can be lowered by an amount $q\,\omega_c$, where $0<q<1$ and $\omega_c$ is the separation between adjacent levels in a cluster. Thus in the quasi-degenerate limit ($\omega_c\ll \omega_0$)  and for small PQD, the clusters remain well separated. Therefore we can confine our considerations 
to the contributions from levels 
deep inside the Fermi cluster to obtain many-body overlap. However, for very large 
clusters ($s_M\to \infty$), 
all clusters are at comparable separations as their constituting levels. In this case, in addition to the Fermi cluster (with its lowest level being somewhat seperated), we have
to take into account 
the neighbouring clusters as well. To incorporate the distinctions
between boundary levels and core levels in the expression of overlap in Eq.~(\ref{OvorgF}), let us use the index $i=l$ and $j=l$
for the lowest levels of filled and empty clusters, respectively. Again, depending on the parent clusters, the lowest levels are at different energy separation from nearest neighbouring cluster. Thus we introduce two more subscripts,  
$s_f$ and $s_e$ respectively, for filled and empty clusters in unperturbed and perturbed eigen-values
in Eq.~(\ref{OvorgF}) as
 \begin{equation}
	|\Delta|^2 = \prod_{i=1}^M\,\prod_{j=M+1}^N\,\left[1-
\frac{2 \delta_{i,s_f}^2}{(\epsilon_{j,s_e}-\epsilon_{i,s_f})^2}+
\frac{\delta_{i,s_f}\,\delta_{j,s_e}}{(\epsilon_{j,s_e}-\epsilon_{i,s_f})^2}\right]\,.
\label{OvorgF2}
\end{equation}  
Here $\delta_{i,s_f}=\epsilon_{i,s_f}-\lambda_{i,s_f}=\delta$ for $i\ne l$ and $\delta_{i=l,s_f}=p D_{s_f-1}$. Similarly,$\delta_{j,s_e}=\epsilon_{j,s_e}-\lambda_{j,s_e}=\delta$ for $i\ne l$ and $\delta_{j=l,s_e}=p D_{s_e-1}$. The clusterwise contributions to the many-body overlap in Eq.~(\ref{OvorgF2}) can be written as
\begin{gather}
	\log |\Delta|^2 = \left( T_0+T_1+T_2+\cdots\right)\,.
\label{Ovsh}\\
\begin{split}\label{T0}
\mbox{Here}\,\,\, T_0 &= - \frac{\delta^2}{\omega_c^2}\, \sum_{j\in s_M} \sum_{\substack{i\in s_M \\ i\ne l}} \frac{1}{(\epsilon_{j,s_M}-\epsilon_{i,s_M})^2}\\
&-pD_{s_M-1}(2p\,D_{s_M-1}-\delta)\sum_{j\in s_M} \frac{1}{(\epsilon_{j,s_M}-\epsilon_{l,s_M})^2}\,,
\end{split}\\
\mbox{with}\,\,\epsilon_{l,s_M}=(\frac{(s_M-1)s_M}{2}+1)\omega_c\,. \nonumber
\end{gather}
\begin{gather}
T_1=T_{1L}+T_{1U}+T_{LU}\,,\label{T1}\\
\begin{split}\label{T1L}
\mbox{with}\,\,\, T_{1L}&=-\delta^2 \sum_{j\in s_M} \sum_{\substack{i\in s_M-1 \\ i\ne l}} \frac{1}{(\epsilon_{j,s_M}-\epsilon_{i,s_M-1})^2}\\
&-p\,D_{s_M-2}(2pD_{s_M-2}-\delta)\sum_{j\in s_M} \frac{1}{(\epsilon_{j,s_M}-\epsilon_{l,s_M-1})^2}\,,
\end{split}\\
\mbox{with}\,\,\epsilon_{l,s_M-1}=(\frac{(s_M-2)(s_M-1)}{2}+1)\omega_c \nonumber
\end{gather}
\begin{gather}
\begin{split}\label{T1U}
T_{1U}&=-\delta^2 \sum_{\substack{j\in s_M+1 \\ j\ne l}}\, \sum_{\substack{i\in s_M \\ i\ne l}} \frac{1}{(\epsilon_{j,s_M+1}-\epsilon_{i,s_M})^2}\\
&-p\,D_{s_M-1}(2p\,D_{s_M-1}-\delta)\sum_{\substack{j\in s_M+1\\j\ne l}} \frac{1}{(\epsilon_{j,s_M+1}-\epsilon_{l,s_M})^2}\\
&-(2\delta^2-\delta p D_{s_M})\sum_{\substack{i\in s_M \\ i\ne l}} \frac{1}{(\epsilon_{l,s_M+1}-\epsilon_{i,s_M})^2}\\
&-\frac{p^2(2D_{s_M-1}^2-D_{s_M}^2)}{(\epsilon_{l,s_M+1}-\epsilon_{l,s_M})^2}\,
\end{split}\\
\mbox{with}\,\,\epsilon_{l,s_M+1}=(\frac{s_M(s_M+1)}{2}+1)\omega_c \nonumber
\end{gather}
\begin{gather}
\begin{split}\label{TLU}
 T_{LU}&=-\delta^2 \sum_{\substack{j\in s_{M+1} \\ j\ne l}} \sum_{\substack{i\in s_{M-1} \\ i\ne l}}\frac{1}{(\epsilon_{j,s_M+1}-\epsilon_{i,s_M-1})^2}\\
&-p\,D_{s_M-2}(2pD_{s_M-2}-\delta)\sum_{\substack{j\in s_M+1 \\j\ne l}} \frac{1}{(\epsilon_{j,s_M+1}-\epsilon_{l,s_M-1})^2}\\
&-(2\delta^2-\delta\,p\,D_{s_M})\sum_{\substack{i\in s_M-1 \\ i\ne l}} \frac{1}{(\epsilon_{l,s_M+1}-\epsilon_{i,s_M-1})^2}\\
&-\frac{p^2(2D_{s_M-2}^2-D_{s_M}^2)}{(\epsilon_{l,s_M+1}-\epsilon_{l,s_M-1})^2}\,
\end{split}
\end{gather}
Here $T_0$ represents the overlap from the main contributing cluster, {\it i.e.} 
the Fermi cluster. The next major contribution, $T_1$, comes
from the neighbouring filled ($s_{M-1}$) and empty ($s_{M+1}$) clusters. 
 Similarly, $T_2$ represents contributions from next nearest neighbour filled
($s_{M-2}$) and empty ($s_{M+2}$) clusters and so on. Upon application of an 
attractive perturbation levels get shifted downwards. The amount of shift increases 
with increasing strength of perturbation. Thus for very weak perturbation, only 
the partially filled cluster participate in the many-body overlap, i.e., 
Eq.~(\ref{Ovsh}) reduces to calculating the first term in $T_0$. As known for 
systems with equi-spaced energy levels, the many-body overlap becomes 
$|\Delta|^2\sim (s_M)^{-\delta^2/\omega_c^2}$, where $s_M$ is the size of
the last partially filled cluster. Now for a half-filled PQD with half-filled 
last cluster, $s_M=\sqrt{N}$. Thus for such a PQD subjected to {\em very weak 
perturbation}, the many-body overlap is given by 
\begin{equation}
|\Delta|^2\sim N^{-\frac{\delta^2}{2\omega_c^2}} \:.
\end{equation}
The phase shift as quoted 
in the main text is $\phi/\pi=\d/\o_c$ for core levels.   
For very large perturbations, all levels from different shells are at
comparable distances (in energy space). In such a case, one can define a 
mean level spacing for the whole system as $d=(s_n-1)\hbar \left( 
\sqrt{(\omega_0^2+\omega_c^2/4)}+\omega_c\right)/N\sim \sqrt{2/N} 
\hbar \omega_0$. In this regime, many-body overlap of PQD with more or
less equi-spaced levels is given by $|\Delta|^2\sim N^{-\delta^2/d^2}$.
In the intermediate regime of perturbation's strength one can not avoid
neighbouring empty and filled shells of the
last (partially) filled shell ($s_M$). In this case Eq.~(\ref{Ovsh}) reduces to 
%\begin{equation}
$\log |\Delta|^2 \sim T_0+T_1\,,\label{lnOv2}$.
%\end{equation}
In the limit of a large cluster ($s_M >>1$) and in the 
quasi-degenerate limit ($\omega_0>>\omega_c$), Eq.~(\ref{lnOv2})
yields the many-boby overlap as a function of cluster size ($s_M$) as
\begin{equation}
        |\Delta|^2 \sim (s_M)^{-\frac{\delta^2}{\omega_c^2}}\,\exp (-\frac{3}{2}p^2\,s_M).\label{OvInt}
\end{equation}
Note that $0<p<1$ and $p$ approaches $1$ with increasing perturbation strength. 
For a half-filled PQD with 
half-filled last cluster under the {\em intermediate regime of 
perturbation} the many-body overlap as a function of $N$ thus becomes
\begin{equation}
|\Delta|^2 \simeq N^{-\frac{\delta^2}{2 \omega_c^2}}
\exp (-\frac{3}{2}p^2\,\sqrt{N})\:.
\end{equation}
It would be worthy to mention 
that in Eq.~(\ref{OvInt}), there are two more exponentially decaying functions 
of $s_M^2$ and $s_M^3$ as $\exp(-a\,s_M^2)$ and $\exp(-b\,s_M^3)$ but 
the multiplicative factor $a, b$ in the argument being very small, their 
contributions are negligible. 
\begin{acknowledgments}
 We thank H. Baranger, E. Mucciolo, G. R\"oder, J. Siewert, S. Tomsovic and D. Ullmo for several helpful discussions. S.B. thanks D. Chaudhuri 
 for many useful discussions, and  MPIPKS, Dresden for financial support, 
where a major portion of this work was done. M.H. thanks the German Research Foundation (DFG) for support through the Emmy-Noether Programme. 

\end{acknowledgments}

%%%%%%%%%%%%%%%%%%%%%%%%%%%%%%%%%%%%%%%%%%%%%%%%%%%%%%%%%%%%%%%%%%%%%%%%%

\end{document}